\documentstyle[twoside,fleqn,npb,epsfig]{article}

\newcommand{\AmS}{{\protect\the\textfont2
  A\kern-.1667em\lower.5ex\hbox{M}\kern-.125emS}}

\newcommand{\be}{\begin{equation}}
\newcommand{\ee}{\end{equation}}
\newcommand{\bea}{\begin{eqnarray}}
\newcommand{\eea}{\end{eqnarray}}
\newcommand{\bdm}{\begin{displaymath}}
\newcommand{\edm}{\end{displaymath}}

\newcommand{\FF}{{\cal F}_{\pi^0\gamma^*\gamma^*}}
\newcommand{\lag}{{\cal L}}
\newcommand{\order}{{\cal O}}
\newcommand{\lapprox}{%
\mathrel{%
\setbox0=\hbox{$<$}
\raise0.6ex\copy0\kern-\wd0
\lower0.65ex\hbox{$\sim$}
}}

\title{
\vspace*{-2.3cm}
\hfill {\normalsize CPT-2002/P.4424} \\[-0.2cm]
\hfill {\normalsize October 2002} \\[0.5cm]
Hadronic light-by-light scattering contribution to $g_\mu
-2$\thanks{Talk presented at the 6th International Symposium on Radiative
Corrections (RADCOR 2002) and the 6th Zeuthen Workshop on Elementary
Particle Theory (Loops and Legs in Quantum Field Theory), Kloster Banz,
Germany, 8-13 September, 2002.
}
}

\author{Andreas Nyf\/feler\address{Centre de Physique Th\'{e}orique,
CNRS-Luminy, Case 907 \\  
F-13288 Marseille Cedex~9, France %\\ 
}%\thanks{}
}

\begin{document}

\begin{abstract}
We briefly review the current status of the hadronic light-by-light
scattering correction to the muon $g-2$. Then we present our
semi-analytical evaluation of the pion-pole contribution, using a
description of the pion-photon-photon transition form factor based on
large-$N_C$ and short-distance properties of QCD. We derive a
two-dimensional integral representation which allows to separate the
generic features from the model dependence, in order to better control
the latter. Finally, we sketch an effective field theory approach to
hadronic light-by-light scattering which yields the leading
logarithmic terms that are enhanced by a factor $N_C$. It also shows
that the modeling of hadronic light-by-light scattering by a
constituent quark loop is not consistent with QCD.
\end{abstract} 

\maketitle

% --------------------------------------------------------------------

\section{Introduction} 

The muon $g-2$ is an important quantity that provides a stringent test
of the Standard Model and which is potentially sensitive to new
physics. However, for this purpose one first needs to well understand
the hadronic contributions, i.e.\ vacuum polarization effects and
light-by-light scattering. 
\vspace*{-0.5cm}
\begin{figure}[h]
\epsfxsize=18pc % will enlarge or reduce the postscript figures based
%on the xsize 
\centerline{\epsfbox{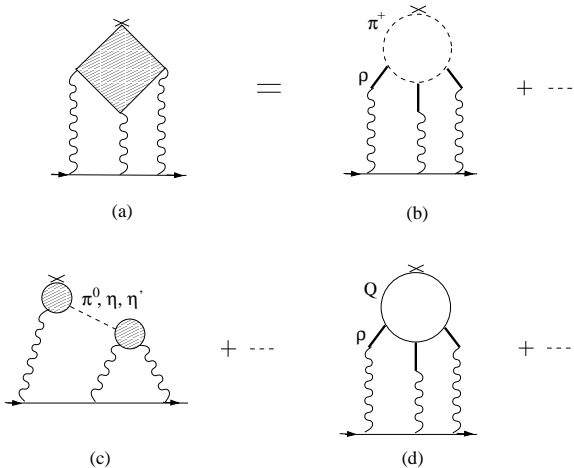}} % postscript image file name
\vspace*{-0.5cm}
\caption{The hadronic light-by-light scattering contribution to the
muon $g-2$. 
}
\label{fig:overview}
\end{figure}
The present picture of hadronic light-by-light scattering is shown in
Fig.~\ref{fig:overview} and the corresponding contributions to $a_\mu$
are listed in Table~\ref{tab:overview}, taking into account the
corrections made in the two full evaluations~\cite{HKS_corr,BPP_corr},
after we had discovered the sign error in the pion-pole
contribution~\cite{KN_pion,a_mu_EFT}.
\vspace*{-0.3cm}
\begin{table}[h]
%\begin{table}[hbt]
%% space before first and after last column: 1.5pc
%% space between columns: 3.0pc (twice the above)
%\setlength{\tabcolsep}{1.5pc}
% -----------------------------------------------------
%% adapted from TeX book, p. 241
%\newlength{\digitwidth} \settowidth{\digitwidth}{\rm 0}
%\catcode`?=\active \def?{\kern\digitwidth}
% -----------------------------------------------------
\caption{Contributions to $a_{{\scriptscriptstyle
\mu}} (\times 10^{10})$ according to Fig.~\ref{fig:overview}. The last
column gives the result when no form factors are used in the couplings
to the photons.} 
\label{tab:overview}   
\begin{center}
\renewcommand{\arraystretch}{1.1}
\begin{tabular}{|l|r@{.}l|r@{.}l|c|c|}
%\begin{tabular*}{\textwidth}{|l|r@{.}l|r@{.}l|r@{.}l|r@{.}l|}
%%\begin{tabular*}{\textwidth}{@{}l@{\extracolsep{\fill}}rrrr}
\hline
Type &
\multicolumn{2}{|c|}{{Ref.~\cite{HKS_corr}}} & 
\multicolumn{2}{|c|}{{Ref.~\cite{BPP_corr}}} & 
Ref.~\cite{KN_pion} & \\ 
\hline  
(b)                & -0 & 5(0.8) & -1 & 9(1.3) & & -4.5 \\ 
(c) & 8 & 3(0.6) & 8 & 5(1.3) & 8.3(1.2)    & $+\infty$ \\
$f_0, a_1$               & 0 & 174$^{\rm a}$ & -0 & 4(0.3) &   
&  \\
(d) & 1 & 0(1.1) & 2 & 1(0.3) &    & $\sim 6$ \\
\hline
Total & 9 & 0(1.5) & 8 & 3(3.2) & 8(4)$^{\rm b}$ &  \\ 
\hline
\end{tabular}
\end{center}
$^{\rm a}$~Only $a_1$ exchange. \\
$^{\rm b}$~Our estimate, using Refs.~\cite{HKS_corr,BPP_corr,KN_pion}. 
\end{table}

There are three classes of contributions to the hadronic four-point
function [Fig.~\ref{fig:overview}(a)], which can be understood from an
effective field theory (EFT) analysis of hadronic light-by-light
scattering~\cite{EdeR_94,a_mu_EFT}: (1) a charged pion loop
[Fig.~\ref{fig:overview}(b)], where the coupling to photons is dressed
by some form factor ($\rho$-meson exchange, e.g.\ via vector meson
dominance (VMD)), (2) the pseudoscalar pole diagrams
[Fig.~\ref{fig:overview}(c)] together with the exchange of heavier
resonances ($f_0, a_1, \ldots$) and, finally, (3) the irreducible part
of the four-point function which was modeled in
Refs.~\cite{HKS_corr,BPP_corr} by a constituent quark loop dressed
again with VMD form factors [Fig.~\ref{fig:overview}(d)]. The latter
can be viewed as a local contribution $\bar\psi
\sigma^{\mu\nu} \psi F_{\mu\nu}$ to $a_\mu$. The two 
groups~\cite{HKS_corr,BPP_corr} used similar, but not identical models
which explains the slightly different results for the dressed charged
pion and the dressed constituent quark loop, although their sum seems
to cancel to a large extent and the final result is essentially given
by the pseudoscalar exchange diagrams. We take the difference of the
results as indication of the error due to the model dependence. 

Our approach to this problem consists of making an ansatz for the
relevant Green's functions in the framework of large-$N_C$ QCD. In
this limit, an infinite set of narrow resonance states contributes in
each channel. Then we perform a matching of the ansatz with chiral
perturbation theory (ChPT) at low energies and with the operator
product expansion (OPE) at high momenta in order to reduce the model
dependence. In practice, it is sufficient to keep a few resonance
states to reproduce the leading behavior in ChPT and the OPE. In this
way we show in Section~\ref{sec:pionpole} that the pseudoscalar
contribution now seems under control, due to our semi-analytical
calculation~\cite{KN_pion}, using a pion-photon-photon form factor
$\FF$ which fulfills the relevant QCD short-distance
constraints~\cite{KN_VAP}, in contrast to the form factors used in
Refs.~\cite{HKS_corr,BPP_corr}.~\footnote{Furthermore, the
calculations in Refs.~\cite{HKS_corr,BPP_corr} were based purely on
numerical approaches.} These findings are also corroborated by an EFT
and large-$N_C$ analysis~\cite{a_mu_EFT} which allows to calculate the
leading and next-to-leading logarithms in
$a_{\mu}^{\mbox{\tiny{LbyL;had}}}$ (Sec.~\ref{sec:EFT}).

% --------------------------------------------------------------------

\section{Pion-pole contribution}
\label{sec:pionpole} 

The contribution from the neutral pion intermediate state is given by
the following two-loop integral (see Ref.~\cite{KN_pion} for all the
details) 
\bea
&&\hspace*{-0.45cm}a_{\mu}^{\mbox{\tiny{LbyL;$\pi^0$}}}= - 
e^6\!\!\int\!\!{d^4 q_1 \over (2\pi)^4}\!\!\!\int\!\!{d^4 q_2 \over
(2\pi)^4} \, \frac{1}{q_1^2 q_2^2 (q_1 + q_2)^2 }\nonumber \\
&&\hspace*{-0.35cm} \times \frac{1}{[(p+ q_1)^2 - m^2][(p - q_2)^2 - m^2]} 
\nonumber \\
&&\hspace*{-0.35cm}\times \left[ {\FF(q_1^2, (q_1 + q_2)^2) \
\FF( q_2^2, 0) \over q_2^2 - M_{\pi^0}^2} \ T_1 \nonumber \right. \\ 
&&\hspace*{-0.25cm}+ \left. {\FF( q_1^2, q_2^2) \ \FF( (q_1 + q_2)^2,
0) \over (q_1 + q_2)^2 - M_{\pi^0}^2} \ T_2 \right], \nonumber \\
&& \label{a_pion_2loop}  
\eea
that involves the convolution of two pion-photon-photon transition
form factors, see Fig.~\ref{fig:overview}(c). The $T_i$ are
polynomials of up to sixth order in the momenta $p,q_1,$ and $q_2$,
with $p^2 = m^2$. 

Since no data on the doubly off-shell form factor $\FF(q_1^2,q_2^2)$
is available, one has to resort to models. In order to proceed with
the analytical evaluation of the two-loop integrals, we considered a
certain class of form factors which includes the ones based on
large-$N_C$ QCD that we studied in Ref.~\cite{KN_VAP}.  For
comparison, we have also used a vector meson dominance (VMD) and a
constant form factor, derived from the Wess-Zumino-Witten (WZW) term.

In large-$N_C$ QCD, the pion-photon-photon form factor is described by 
a sum over an infinite set of narrow vector resonances, involving
arbitrary couplings, although there are constraints at long and short
distances. The normalization is given by the WZW term, $\FF(0,0) = -
N_C / (12 \pi^2 F_\pi)$, whereas the OPE tells us that
\bea
&& \lim_{\lambda\to \infty}\,\FF(\lambda^2 q^2, (p-\lambda
q)^2)  \nonumber \\  
&& = \frac{2}{3}\,\frac{F_\pi}{q^2}\,\bigg\{ \frac{1}{\lambda^2}\, 
+ \frac{1}{\lambda^3} \frac{q \cdot p}{q^2} 
+\,{\cal O}\left({1 \over \lambda^4}\right)\bigg\}\,. 
\label{OPE_FF}
\eea
In the following, we consider the form factors that are obtained by
truncation of the infinite sum in large-$N_C$ QCD to one (lowest meson
dominance, LMD), and two (LMD+V), vector resonances per
channel, respectively:
\be
\FF^{\mbox{{\tiny LMD}}}(q_1^2,q_2^2) =   \!{F_\pi \over 3} {
q_1^2 + q_2^2 - c_V  \over (q_1^2 - M_V^2) (q_2^2 - M_V^2) }, 
\label{FF_LMD}
\ee
\bea
&&\hspace*{-0.7cm}\FF^{\mbox{{\tiny LMD+V}}}(q_1^2,q_2^2)\!=\!{F_\pi
\over 3} \Bigg\{\!\bigg(\!\!q_1^2\!q_2^2
(q_1^2\!+\!q_2^2)\!+\!h_1\!(q_1^2\!+\!q_2^2)^2 
\nonumber \\
&&\!\!+ h_2q_1^2 q_2^2 + h_5 (q_1^2\!+\!q_2^2) + 
h_7 \bigg) \bigg/ \bigg( (q_1^2 - M_{V_1}^2) \nonumber \\
&& \times (q_1^2 - M_{V_2}^2) (q_2^2 - M_{V_1}^2) (q_2^2 -
M_{V_2}^2) \bigg) \Bigg\}, \label{FF_LMD+V} 
\eea
with the constants $c_V \,= \, N_C M_V^4 / (4\pi^2 F_\pi^2)$ and $h_7
\,= \, - N_C M_{V_1}^4 M_{V_2}^4 / (4\pi^2 F_\pi^2)$.  The parameters
$h_1,h_2,$ and $h_5$ in the LMD+V form factor are not fixed by the
normalization and the leading term in the OPE. We have determined
these coefficients phenomenologically~\cite{KN_VAP,KN_pion}.  In
particular, $\FF(\!-Q^2,\!0)$ with one photon on-shell behaves like
$1/Q^2$ for large spacelike momenta, $Q^2\!=\!-q^2$. Whereas the LMD
form factor does not have such a behavior, it can be reproduced with
the LMD+V ansatz, provided that $h_1\!=\!0$. A fit to the data yields
moreover $h_5\!=\!6.93 \pm 0.26~\mbox{GeV}^4$. Analyzing the
experimental data for the decay $\pi^0\!\to\!e^+ e^-$ leads to the
loose bound $|h_2|\!\lapprox\!20~\mbox{GeV}^2$.

Note that the usual VMD form factor $\FF^{\mbox{\tiny
VMD}}(q_1^2,q_2^2) \sim 1/ [(q_1^2 - M_V^2) (q_2^2 - M_V^2)]$ does
{\it not} correctly reproduce the OPE in Eq.~(\ref{OPE_FF}).

For the form factors discussed above one can perform {\it all} angular
integrations in the two-loop integrals analytically~\cite{KN_pion}
using the method of Gegenbauer polynomials~\cite{Gegenbauer}. The key
observation is that all form factors can be written as follows 
\be
\FF(q_1^2, q_2^2) = \tilde f(q_1^2) - \sum_{M_{V_i}} 
{\tilde g_{M_{V_i}}(q_1^2) \over q_2^2 - M_{V_i}^2}. 
\ee
This allows to cancel all dependences on $q_1\!\cdot\!q_2$ in the
numerators in $\FF(q_1^2\!,\!(q_1\!+\!q_2)^2)$ in the loop integrals
in Eq.~(\ref{a_pion_2loop}). Then one writes the propagators as (for
Euclidean momenta $K,L$)  
\be
{1 \over (K\!-\!L)^2\!+\!M^2} = {Z_{KL}^M \over |K| |L|}
\sum_{n=0}^\infty (Z_{KL}^M)^n C_n(\hat K \cdot \hat L), 
\ee
with $Z_{KL}^M = (K^2\!+\!L^2\!+\!M^2 - [(K^2\!+\!L^2\!+\!M^2)^2\!-\!4K^2
L^2]^{1/2}) / (2 |K| |L|)$, and uses the orthogonality
properties of the Gegenbauer polynomials  
\bea
\lefteqn{ \int d\Omega(\hat K) \, C_n(\hat Q_1 \cdot \hat K) \, 
C_m(\hat K \cdot \hat Q_2) } \nonumber \\
\hspace*{1cm}& = & 2 \pi^2 {\delta_{nm} \over n + 1} \, 
C_n(\hat Q_1 \cdot \hat Q_2), 
\eea
where for instance $\hat Q_1 \cdot \hat K$ denotes the cosine of the
angle between the four-dimensional vectors $Q_1$ and $K$. After
performing the angular integrations in this way, the pion-exchange
contribution to $a_\mu$ can be written as a two-dimensional integral
representation, where the integration runs over the moduli of the
Euclidean momenta
\bea
a_{\mu}^{\mbox{\tiny{LbyL;$\pi^0$}}} & = & \int_0^\infty d Q_1 
\int_0^\infty d Q_2 \nonumber \\
&& \quad \times \sum_i w_i(Q_1,Q_2) \ f_i(Q_1,Q_2) ,  
\eea
with universal [for the above class of form factors] weight functions
$w_i$ (rational functions, square roots and
logarithms)~\cite{KN_pion}. The dependence on the form factors resides
in $f_i$. In this way we could separate the generic features of the
pion-pole contribution from the model dependence and thus better
control the latter. This is not possible anymore in the final
analytical result (as a series expansion) for
$a_{\mu}^{\mbox{\tiny{LbyL;$\pi^0$}}}$ derived in
Ref.~\cite{Blokland_etal}.  Note that the analytical result has not
the same status here as for instance in QED. One has to keep in mind
that there is an intrinsic uncertainty in the form factor of $10 -
30~\%$, furthermore the VMD form factor used in that reference has the
wrong high-energy behavior.

The weight functions $w_i$ in the main contribution are positive and
peaked around momenta of the order of $0.5~\mbox{GeV}$. There
is, however, a tail in one of these functions, which produces for the
constant WZW form factor a divergence of the form $\ln^2\!\Lambda$
for some UV-cutoff $\Lambda$. Other weight
functions have positive and negative contributions in the low-energy
region, which lead to a strong cancellation in the corresponding
integrals.

In Table~\ref{tab:api_models} we present the numerical results for the
different form factors.  
\begin{table}[b]
%% space before first and after last column: 1.5pc
%% space between columns: 3.0pc (twice the above)
%\setlength{\tabcolsep}{1.5pc}
% -----------------------------------------------------
%% adapted from TeX book, p. 241
%\newlength{\digitwidth} \settowidth{\digitwidth}{\rm 0}
%\catcode`?=\active \def?{\kern\digitwidth}
% -----------------------------------------------------
\caption{Results for 
$a_{{\scriptscriptstyle \mu}}^{\mbox{\tiny{LbyL;$\pi^0$}}}$ for the
different form factors. In the WZW model we used a cutoff of
$1~\mbox{GeV}$ in the divergent contribution. 
}
\label{tab:api_models}   
\begin{center}
\renewcommand{\arraystretch}{1.1}
\begin{tabular}{|l|r@{.}l|}
%%\begin{tabular*}{\textwidth}{@{}l@{\extracolsep{\fill}}rrrr}
\hline
Form factor &
\multicolumn{2}{|c|}{{$a_{{\scriptscriptstyle
\mu}}^{\mbox{\tiny{LbyL;$\pi^0$}}} \times 10^{10}$}}    
\\ 
\hline  
WZW     & \hspace*{0.5cm} 12 & 2 \\  
VMD     &  5 & 6 \\ 
LMD     & 7 & 3 \\
LMD+V ($h_2 = 0~\mbox{GeV}^2$)
        & 5 & 8 \\
\hline
\end{tabular}
\end{center}
\end{table}
All form factors lead to very similar results (apart from
WZW). Judging from the shape of the weight functions described above,
it seems more important to correctly reproduce the slope of the form
factor at the origin and the available data at intermediate
energies. On the other hand, the asymptotic behavior at large $Q_i$
seems not very relevant.  The results for the LMD+V form factor are
rather stable under the variation of the parameters, except for
$h_2$. If all other parameters are kept fixed, our result changes in
the range $|h_2| < 20~\mbox{GeV}^2$ by $\pm 0.9 \times 10^{-10}$ from
the value for $h_2 = 0$.

Thus, with the LMD+V form factor, we get 
\be 
a_{\mu}^{\mbox{\tiny{LbyL;$\pi^0$}}} = + 5.8~(1.0) \times 10^{-10}
\, , 
\ee
where the error includes the variation of the parameters and the
intrinsic model dependence. A similar short-distance analysis in the
framework of large-$N_C$ QCD and including quark mass corrections for
the form factors for the $\eta$ and $\eta^\prime$ was beyond the scope
of Ref.~\cite{KN_pion}. We therefore used VMD form factors fitted to
the available data for $\FF(-Q^2,0)$ to obtain our final estimate 
\bea
a_{\mu}^{\mbox{\tiny{LbyL;PS}}} & \equiv & 
a_{\mu}^{\mbox{\tiny{LbyL;$\pi^0$}}}
+ a_{\mu}^{\mbox{\tiny{LbyL;$\eta$}}}\vert_{\mbox{\tiny VMD}} +
a_{\mu}^{\mbox{\tiny{LbyL;$\eta^\prime$}}}\vert_{\mbox{\tiny VMD}}
\nonumber \\
& = & + 8.3~(1.2) \times 10^{-10} \, . 
\eea
An error of 15~\% for the pseudoscalar pole contribution seems
reasonable, since we impose many theoretical constraints from long 
and short distances on the form factors. Furthermore, we use
experimental information whenever available. 

% --------------------------------------------------------------------

\section{EFT approach to $a_{\mu}^{\mbox{\tiny{LbyL;had}}}$}
\label{sec:EFT}

In Ref.~\cite{a_mu_EFT} we discussed an EFT approach to hadronic
light-by-light scattering based on an effective Lagrangian that
describes the physics of the Standard Model well below 1~GeV. It
includes photons, light leptons, and the pseudoscalar mesons and obeys
chiral symmetry and $U(1)$ gauge invariance.

The leading contribution to $a_{\mu}^{\mbox{\tiny{LbyL;had}}}$, of
order $p^6$, is given by a finite loop of charged pions with
point-like electromagnetic vertices, see Fig.~\ref{fig:overview}(b).
Since this contribution involves a loop of hadrons, it is subleading
in the large-$N_C$ expansion.

\begin{figure}[h]
\epsfxsize=18pc % will enlarge or reduce the postscript figures based
%on the xsize 
\centerline{\epsfbox{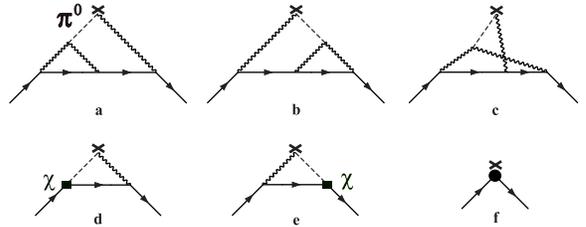}} % postscript image file name
\caption{The graphs contributing to
$a_{\mu}^{\mbox{\tiny{LbyL;$\pi^0$}}}$ at lowest order in the effective
field theory. 
\label{fig:EFT}}
\end{figure}

At order $p^8$ and at leading order in $N_C$, we encounter the
divergent pion-pole contribution, diagrams (a) and (b) of
Fig.~\ref{fig:EFT}, involving two WZW vertices.  
The diagram (c) is actually finite. The divergences of the triangular
subgraphs in the diagrams (a) and (b) are removed by inserting the
counterterm $\chi$ from the Lagrangian $\lag^{(6)} = (\alpha^2 / 4
\pi^2 F_0) \ \chi \ {\overline\psi} \gamma_\mu \gamma_5 \psi \,
\partial^\mu \pi^0 + \cdots$, see the one-loop diagrams (d) and
(e). Finally, there is an overall divergence of the two-loop diagrams
(a) and (b) that is removed by a local counterterm, diagram (f).
Since the EFT involves such a local contribution, we will not be able
to give a precise numerical prediction for
$a_{\mu}^{\mbox{\tiny{LbyL;had}}}$.

Nevertheless, it is interesting to consider the leading and
next-to-leading logarithms that are in addition enhanced by a factor
$N_C$ and which can be calculated using the renormalization
group~\cite{a_mu_EFT}. The EFT and large-$N_C$ analysis tells us that 
\bea
&&\!\!\!\!\!\!\!\!a_{\mu}^{\mbox{\tiny{LbyL;had}}} = 
\left( {\alpha \over \pi} \right)^3  \Bigg\{
f\left({M_{\pi^\pm} \over m_\mu}, {M_{K^\pm} \over m_\mu}\right)
 \nonumber \\
&&\!\!\!\!\!\!\!\!\!+ N_C \left( {m_\mu^2 \over 16 \pi^2
F_\pi^2} {N_C \over 3} \right)
\left[ \ln^2 {\mu_0 \over m_\mu} + c_1 \ln {\mu_0 \over m_\mu} + c_0
\right]  \nonumber \\ 
&&\!\!\!\!\!\!\!\!\!+ \order \left(\!{m_\mu^2 \over \mu_0^2} \times
\mbox{log's}\!\right) + \order \left(\!{m_\mu^4 \over \mu_0^4} 
N_C \times \mbox{log's}\!\right)\!\!\Bigg\},  \label{a_mu_EFT_N_C}
\eea
where $f(M_{\pi^\pm}\!/\!m_\mu,\!M_{K^\pm}\!/\!m_\mu)\!=\!\!-0.038$
represents the charged pion and kaon-loop that is formally of order
one in the chiral and $N_C$ counting and $\mu_0$ denotes some hadronic
scale, e.g.\ $M_\rho$.  The coefficient ${\cal C}$ of the log-square
term in the second line is universal and of order $N_C$, since
$F_\pi\!=\!\order(\sqrt{N_C})$.  

Unfortunately, although the logarithm is sizeable, in
$a_{\mu}^{\mbox{\tiny{LbyL;$\pi^0$}}}$ there occurs a cancellation
between the log-square and the log-term. If we fit our result for the
VMD form factor for large $M_\rho$ to an expression as given in
Eq.~(\ref{a_mu_EFT_N_C}), we obtain
\bea
a_{\scriptscriptstyle{\mu;\mbox{\tiny{VMD}}}}^{\mbox{\tiny{LbyL}}; \pi^0}
& \doteq &\!\!\left( {\alpha \over \pi} \right)^3 {\cal C} 
\ \ \left[ \ln^2 {M_\rho \over m_\mu} + c_1 \ln {M_\rho \over m_\mu} + c_0
\right]  \nonumber \\
& \stackrel{\mbox{\tiny{Fit}}}{=} &\!\!\left( {\alpha \over  \pi}
\right)^3 {\cal C}  
\ \ \left[ 3.94 - 3.30 + 1.08 \right] \nonumber \\
& = & \hspace*{0cm}\!\!\left[ 12.3 - 10.3 + 3.4 \right] \times 10^{-10}
\nonumber \\
& = &\!\!5.4 \times 10^{-10} \, , 
\eea
which is confirmed by the analytical result of
Ref.~\cite{Blokland_etal} (setting for simplicity $M_{\pi^0} =
m_\mu$): $a_{\scriptscriptstyle{\mu;
\mbox{\tiny{VMD}}}}^{\mbox{\tiny{LbyL}}; \pi^0} = [12 - 8.0 + 1.7]
\times 10^{-10} = 5.7 \times 10^{-10}$. This 
cancellation is now also visible in the revised version of
Ref.~\cite{Ramsey-Musolf_Wise}. In that paper the remaining parts of
$c_1$ have been calculated: $c_1 = - 2 \chi(\mu_0) / 3 + 0.237 =
-0.93^{+0.67}_{-0.83}$, with our conventions for $\chi$ and
$\chi(M_\rho)_{{\rm exp}} = 1.75^{+1.25}_{-1.00}$.

Finally, the EFT analysis shows that the modeling of hadronic
light-by-light scattering by a constituent quark loop is not
consistent with QCD. The latter has a priori nothing to do with the
full quark loop in QCD which is dual to the corresponding contribution
in terms of hadronic degrees of freedom.
Equation~(\ref{a_mu_EFT_N_C}) tells us that at leading order in $N_C$
any model of QCD has to show the behavior
$a_\mu^{\mbox{\tiny{LbyL;had}}}
\sim (\alpha/\pi)^3 N_C [N_C m_\mu^2 / (48 \pi^2 F_\pi^2)]
\ln^2\Lambda$, with a universal coefficient, if one sends the cutoff
$\Lambda$ to infinity. From the analytical result given in
Ref.~\cite{Laporta_Remiddi} for the quark loop, one obtains the
behavior $a_\mu^{\mbox{\tiny LbyL;CQM}} \sim (\alpha/\pi)^3 N_C
(m_\mu^2 / M_Q^2) + \ldots$, for $M_Q \gg m_\mu$, if we interpret the
constituent quark mass $M_Q$ as a hadronic cutoff.  Even though one
may argue that $N_C / 48 \pi^2 F_\pi^2$ can be replaced by $1/M_Q^2$,
the log-square term is not correctly reproduced with this
model. Therefore, the constituent quark model (CQM) cannot serve as a
reliable description for the dominant contribution to
$a_\mu^{\mbox{\tiny{LbyL;had}}}$, in particular, its sign.  Moreover,
we note that the pion-pole contribution is {\it infrared finite} in
the chiral limit\footnote{This can be shown by studying the low
momentum behavior of the weight functions $w_i$ corresponding to the
two-loop \nopagebreak[1] diagrams 2(a)--(c) and the one-loop diagrams
2(d)+(e) (given in Ref.~\cite{a_mu_EFT}) for $M_{\pi^0} \to 0$, see
also Ref.~\cite{Ramsey-Musolf_Wise}.}, whereas the quark loop shows an
infrared divergence $\ln(M_Q/m_\mu)$ for $M_Q \to
0$~\cite{Laporta_Remiddi}.

% --------------------------------------------------------------------

\section{Conclusions}

The pseudoscalar pole con\-tri\-bu\-tion
$a_{\mu}^{\mbox{\tiny{LbyL;PS}}}$ seems to be under control at the
15~\% level. Moreover, the EFT and large-$N_C$ analysis provides a
systematic approach to $a_{\mu}^{\mbox{\tiny{LbyL;had}}}$ and yields
the leading and next-to-leading logarithmic terms, enhanced by a
factor $N_C$.  It also shows that the modeling of hadronic
light-by-light scattering by a constituent quark loop is not
consistent with QCD.  Since model calculations for the dressed charged
pion and the dressed constituent quark loop yield slightly different
results, our present estimate reads (by adding the errors linearly)
\be
a_{\mu}^{\mbox{\tiny{LbyL;had}}} = +\,8~(4) \times 10^{-10} \, . 
\ee

% --------------------------------------------------------------------

\section*{Acknowledgments}
I am grateful to my collaborators M.\ Knecht, M.\ Perrottet and E.\ de
Rafael. I thank the organizers of RADCOR / Loops \& Legs 2002 for
providing a pleasant atmosphere and the Schweizerischer
Na\-tio\-nal\-fonds for financial support.

% --------------------------------------------------------------------


\begin{thebibliography}{99}

\bibitem{HKS_corr}
M.\ Hayakawa and T.\ Kinoshita, Phys.\ Rev.\ D {\bf 66}, 019902 (E)
(2002) and hep-ph/0112102.
%M.\ Hayakawa and T.\ Kinoshita, 
%%``Comment on the sign of the pseudoscalar pole contribution to the
%%muon g-2,'' 
%hep-ph/0112102
%%%CITATION = HEP-PH 0112102;%%

\bibitem{BPP_corr}
J.\ Bijnens, E.\ Pallante, and J.\ Prades, Nucl.\ Phys.\ {\bf B626},
410 (2002). 
%hep-ph/0112255. 

\bibitem{KN_pion}
M.\ Knecht and A.\ Nyf\/feler, Phys.\ Rev.\ D {\bf 65}, 073034 (2002). 
%hep-ph/0111058 

\bibitem{a_mu_EFT} 
M.\ Knecht, A.\ Nyf\/feler, M.\ Perrottet, and E.\ de Rafael, Phys.\ Rev.\
Lett.\ {\bf 88}, 071802 (2002).  
% hep-ph/0111059  

\bibitem{EdeR_94}
E.\ de Rafael, Phys. Lett. B {\bf 322}, 239 (1994).

\bibitem{KN_VAP}
M.\ Knecht and A.\ Nyf\/feler, %{\it Resonance estimates of
%$\order(p^6)$ low-energy constants and QCD short-distance constraints}, 
Eur. Phys. J. C {\bf 21}, 659 (2001).

\bibitem{Gegenbauer}
%J.L.\ Rosner, %{\it Higher-Order Contributions to the Divergent Part of
%%$Z_3$ in a Model Quantum Electrodynamics}, 
%Ann.\ Phys.\ (N.Y.) {\bf 44}, 11 (1967); 
M.J.\ Levine and R.\ Roskies, %{\it Hyperspherical approach to quantum
%electrodynamics: sixth-order magnetic moment}, 
Phys.\ Rev.\ D {\bf 9},
421 (1974); M.J.\ Levine, E.\ Remiddi, and R.\ Roskies, 
%{\it Analytic contributions to the $g$ factor of the electron in
%sixth order}, 
{\it ibid.} {\bf 20}, 2068 (1979).

\bibitem{Blokland_etal}
I.\ Blokland, A.\ Czarnecki, and K.\ Melnikov, Phys.\ Rev.\ Lett.\ {\bf
88}, 071803 (2002). 
%``Pion pole contribution to hadronic light-by-light scattering and
%muon anomalous magnetic moment,'' 
%hep-ph/0112117.
%%CITATION = HEP-PH 0112117;%%

\bibitem{Ramsey-Musolf_Wise} 
M.\ Ramsey-Musolf and M.B.\ Wise, Phys.\ Rev.\ Lett.\ {\bf 89}, 041601 
(2002). 
%``Hadronic light-by-light contribution to muon g-2 in chiral
% perturbation  theory,'' 
%hep-ph/0201297.
%%CITATION = HEP-PH 0201297;%%

\bibitem{Laporta_Remiddi} 
S.\ Laporta and E.\ Remiddi,
%``The Analytical value of the electron light-light graphs
% contribution to the muon (g-2) in QED,'' 
Phys.\ Lett.\ B {\bf 301}, 440 (1993). 
%%CITATION = PHLTA,B301,440;%%

\end{thebibliography}
\end{document}